# Mutual Intolerance Effect in Entangled Qubit Pairs

Moses Fayngold

*Department of Physics, New Jersey Institute of Technology, Newark, NJ 07102*

Analysis of the recently proposed thought experiment with the path-entangled photon pairs is extended here to spin-entangled electron pairs. The detailed comparison of the two cases showed the range of distinctions and similarities in their monitoring. The general results contradict the "Concurrency Rule" stating that intimately linked characteristics of a simple system must change concurrently under changing conditions. Instead, the analysis showed that the system's coherence, while changing continuously with entanglement strength on the global scale, remains zero on the local level. This effect, common for bi-photons and bi-fermions, can be named "total mutual intolerance" between local and global coherence. We can thus predict mutual intolerance as a general effect for all pairs of entangled qubits regardless of their physical nature.

## 1. Spin state in different bases

Elimination of local interference in the path-entangled photon pairs was first described in [1, 2]. But the analysis was restricted only to the case with maximally strong entanglement. Studying the effect in *all domain* of the entanglement strength (ES) was the topic of our work [3]. Here we extend our thought experiment with a bi-photon proposed in [3] to a spin-entangled bi-fermion with spin $S=1/2$. To compare both cases, we reproduce here the experimental setup in [3] for a bi-photon created by source S (Fig.1).

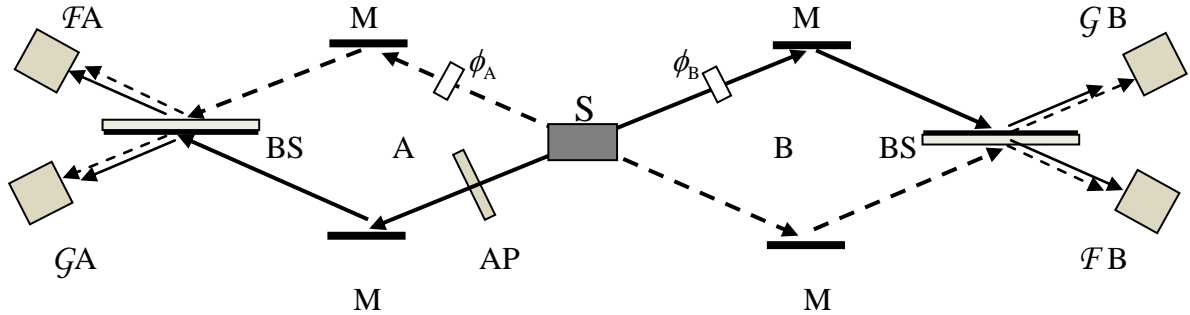

**Fig. 1** The setup for a thought experiment proposed in [3].
The inserted plate AP monitors the ratio of superposition amplitudes of paths **1**(the solid line) and **2** (the dashed line). Mirrors M reflect the paths directly to the respective beam splitters (BS); $\phi_A$, $\phi_B$ are the phase shifters; $\mathcal{F}A$, $\mathcal{G}A$, $\mathcal{F}B$, $\mathcal{G}B$ are photon detectors.

The paths **1** and **2** in Fig.1, making angle $\beta$ in our physical space $V$ are mutually orthogonal eigenstates in the system's Hilbert space $\mathcal{H}$. While $\mathbf{1}\cdot\mathbf{2}=\cos\beta$ in $V$, we have $\langle\mathbf{1}||\mathbf{2}\rangle=0$ in $\mathcal{H}$.

Turning to fermions, we describe in this section monitoring a single qubit state. The term qubit applies to spin ½ fermions, with two eigenstates each, like in the bi-photon case in [1-3]. On the way we will point at similarities and differences between an electron qubit and two paths photon qubit.

An arbitrary spin state $|s\rangle$ is usually represented as an arrow **s** in the Bloch sphere (Fig.2)

$$|s\rangle = a|\uparrow\rangle + b e^{i\varphi}|\downarrow\rangle; \qquad a = \cos\frac{\theta}{2}, \quad b = \sin\frac{\theta}{2} \qquad (1.1)$$

(the used notations imply $a$ and $b$ being real positive without loss of generality). Its "antipode" $|\bar{s}\rangle$ is an arrow $\bar{\mathbf{s}} \equiv -\mathbf{s}$ with $\bar{\theta} = \pi - \theta$, $\bar{\varphi} = \varphi + \pi$:

$$|\bar{s}\rangle = b|\uparrow\rangle - a e^{i\varphi}|\downarrow\rangle \qquad (1.2)$$

Angles $\varphi$, $\theta$ can be monitored by varying the direction of magnetic field **B** in a Stern-Gerlach device used for creating state $|s\rangle$. So $\theta$, $\varphi$ in (1.1) actually indicate the direction of **B**.

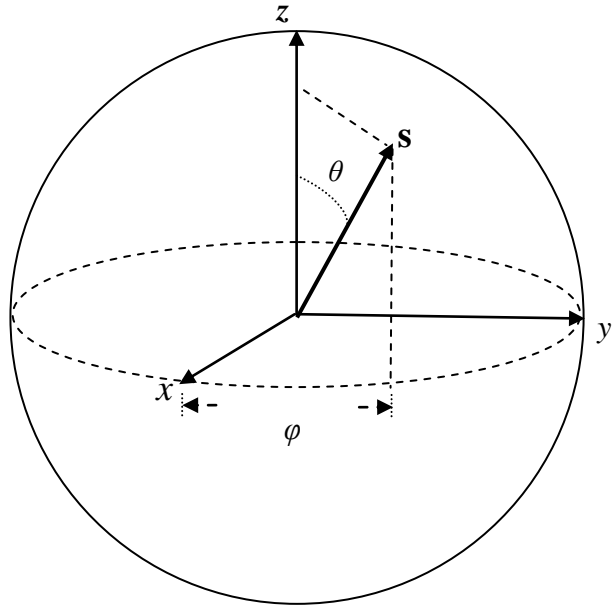

**Fig. 2**
Graphical representation of $|s\rangle$ in the Bloch sphere with the polar angle $\theta$ and azimuth $\varphi$

While $\mathbf{s} \cdot \bar{\mathbf{s}} = -1$ in $V$, we have $\langle \mathbf{s} \| \bar{\mathbf{s}} \rangle = 0$ in $\mathcal{H}$.

Actually, representing spin states as vectors $\mathbf{s}$, $\bar{\mathbf{s}}$ in $V$ is largely misleading [4, 5]. It contradicts indeterminacy relations for spin components. While geometrical projection of $\mathbf{s}$ onto the z-axis represents its $s_z$ component $s_z = \cos\theta$, its von Neumann projection represents respective probability amplitude $\cos\frac{\theta}{2}$. With all that, the crude visualization as in Fig.1 is still helpful when we describe monitoring fermion states in terms of $\varphi, \theta$. It also helps when comparing photon and fermion states. For instance, while a photon requires installing the AP at least in one of the paths to monitor its superposition amplitudes, the fermion states can be monitored just by changing $\varphi, \theta$. Inserting AP in case [3] changes the system's Hamiltonian and expands $\mathcal{H}$ from 2D to 3D due to additional eigenstate $|0\rangle$ - photon absorption. In contrast, changing $\varphi, \theta$ to monitor spin states is just a rotation of basis in $\mathcal{H}$.

With all these distinctions, the corresponding math is identical in both cases.

Like in case [3], we can write (1.1, 2) in the matrix form

$$\begin{pmatrix} |\mathbf{s}\rangle \\ |\bar{\mathbf{s}}\rangle \end{pmatrix} = \mathcal{R}_\mathbf{s} \begin{pmatrix} |\uparrow\rangle \\ |\downarrow\rangle \end{pmatrix}, \qquad \mathcal{R}_\mathbf{s} \equiv \begin{pmatrix} a & be^{i\varphi} \\ b & -ae^{i\varphi} \end{pmatrix} \qquad (1.3)$$

In view of (1.1), the determinant of $\mathcal{R}_\mathbf{s}$ is $D(\mathcal{R}_\mathbf{s}) = -e^{i\varphi}$.

There are two equivalent ways of changing the amplitudes in state (1.1). We can physically change the state by passing it through a differently oriented field $\mathbf{B}'$. Alternatively, we can change the basis by switching from $(x, y, z)$ triad to some other reference axis $\mathbf{e}$.

Suppose we do switch to the **e**-basis characterized by the polar angle $\chi$ and azimuth $\delta$ with respect to the initial system $(x, y, z)$. The respective eigenstates $|\mathbf{e}\rangle$ and $|\bar{\mathbf{e}}\rangle$ will be

$$\left. \begin{array}{l} |\mathbf{e}\rangle = m|\uparrow\rangle + ne^{i\delta}|\downarrow\rangle \\ |\bar{\mathbf{e}}\rangle = n|\uparrow\rangle - me^{i\delta}|\downarrow\rangle \end{array} \right\}, \quad m = \cos\frac{\chi}{2}, \; n = \sin\frac{\chi}{2} \qquad (1.8)$$

or

$$\begin{pmatrix} |\mathbf{e}\rangle \\ |\bar{\mathbf{e}}\rangle \end{pmatrix} = \mathcal{R}_\mathbf{e} \begin{pmatrix} |\uparrow\rangle \\ |\downarrow\rangle \end{pmatrix}, \; \mathcal{R}_\mathbf{e} \equiv \begin{pmatrix} m & ne^{i\delta} \\ n & -me^{i\delta} \end{pmatrix}, \; D(\mathcal{R}_\mathbf{e}) = -e^{i\delta} \qquad (1.9)$$

The inverse transformation is

$$\begin{pmatrix} |\uparrow\rangle \\ |\downarrow\rangle \end{pmatrix} = \mathcal{R}_\mathbf{e}^{-1} \begin{pmatrix} |\mathbf{e}\rangle \\ |\bar{\mathbf{e}}\rangle \end{pmatrix}, \qquad \mathcal{R}_\mathbf{e}^{-1} \equiv \begin{pmatrix} m & n \\ ne^{-i\delta} & -me^{-i\delta} \end{pmatrix} = \mathcal{R}_\mathbf{e}^\dagger \qquad (1.10)$$

Combining (1.3) and (1.10) and denoting $\varphi - \delta \equiv \eta$ represents $|\mathbf{s}\rangle, |\bar{\mathbf{s}}\rangle$ in the new basis:

$$\begin{pmatrix} |\mathbf{s}\rangle \\ |\bar{\mathbf{s}}\rangle \end{pmatrix} = \mathcal{R}_\mathbf{s}^\mathbf{e} \begin{pmatrix} |\mathbf{e}\rangle \\ |\bar{\mathbf{e}}\rangle \end{pmatrix}, \quad \mathcal{R}_\mathbf{s}^\mathbf{e} \equiv \mathcal{R}_\mathbf{s}\, \mathcal{R}_\mathbf{e}^{-1} = \begin{pmatrix} u & v \\ -e^{i\eta}v^* & e^{i\eta}u^* \end{pmatrix} \quad (1.11)$$

or

$$\begin{aligned} |\mathbf{s}\rangle &= u|\mathbf{e}\rangle + v|\bar{\mathbf{e}}\rangle \\ |\bar{\mathbf{s}}\rangle &= -e^{i\eta}\left(v^*|\mathbf{e}\rangle - u^*|\bar{\mathbf{e}}\rangle\right) \end{aligned}, \quad \begin{aligned} u &\equiv am + bn\, e^{i\eta} \\ v &\equiv an - bm\, e^{i\eta} \end{aligned} \quad (1.12)$$

Transformation (1.11, 12) cannot be expressed in terms of a *real* angle $\gamma$ between **s** and **e**. An attempt to represent $u$, $v$ by analogy with (1.1, 10) in terms of $\gamma$ will give, according to (1.12)

$$\begin{aligned} u &= \cos\frac{\gamma}{2} = \cos\frac{\theta}{2}\cos\frac{\chi}{2} + \sin\frac{\theta}{2}\sin\frac{\chi}{2}\, e^{i\eta}, \\ v &= \sin\frac{\gamma}{2} = \cos\frac{\theta}{2}\sin\frac{\chi}{2} - \sin\frac{\theta}{2}\cos\frac{\chi}{2}\, e^{i\eta} \end{aligned} \quad (1.13)$$

We see that **s** and **e** generally do not make a real angle with each other. This is another illustration of inadequacy of the vector image of **s** and another snag in mapping spin states from $\mathcal{H}$ onto *V*. But the probabilities of the outcomes $|\mathbf{e}\rangle$ or $|\bar{\mathbf{e}}\rangle$ in measurements of state $|\mathbf{s}\rangle$ in the **e**-basis are real functions of $\theta$, $\chi$ and $\eta$:

$$\begin{aligned} \mathcal{P}(\mathbf{e}) &= a^2 m^2 + b^2 n^2 + 2abmn\,\cos\eta, \\ \mathcal{P}(\bar{\mathbf{e}}) &= a^2 n^2 + b^2 m^2 - 2abmn\,\cos\eta \end{aligned} \quad \mathcal{P}(\mathbf{e}) + \mathcal{P}(\bar{\mathbf{e}}) = 1 \quad (1.14)$$

The periodic dependence on $\eta$ indicates particle interference with itself similar to the photon self-interference in a Mach-Zehnder interferometer. The ratio of the amplitude of the periodic term to the constant term gives the measure of visibility (contrast) $\mathcal{V}$ of the interference pattern:

$$\mathcal{V}_\mathbf{e} = \frac{2abmn}{a^2 m^2 + b^2 n^2}, \qquad \mathcal{V}_{\bar{\mathbf{e}}} = \frac{2abmn}{a^2 n^2 + b^2 m^2} \quad (1.15)$$

The bi-photon experiment shown in Fig.1 could be modified by removing AP and inserting a screen near S with two narrow perforations that would make it like a double-slit experiment. The distance between perforations must greatly exceed the photon wavelength. Then we could select different paths from the respective spherical waves, e.g., one path from the first perforation perpendicular to the screen and one from the second perforation making some other angle and thereby having weaker amplitude. Or else, perforations themselves could have different size. But such setups would be even more complicated than the one with AP.

We could similarly complicate the bi-fermion experiment by inserting in the electron's way a plate partially absorbing only one eigenstate, say, $|\downarrow\rangle$, but not the other.

Overall, we see wide overlap of similar operations in both cases, which may explain, at least partially, identical math describing manipulations of their states.

## 2. Composite qubit states

Here we consider composite qubit states (bi-qubits). Two *disentangled* qubits A and B can be represented by vectors **s**, **s**′. Assume their spins have opposite $z$-projections. Then their polar angles $\theta$, $\theta'$ are related as $\theta' = \pi - \theta$, while $\varphi$, $\varphi'$, and thereby $\phi \equiv \varphi - \varphi'$ may be arbitrary.

Now turn to entangled states. A commonly used case is a pair of mutually receding electrons A and B, each with eigenstates $|\uparrow\rangle$ and $|\downarrow\rangle$ in the $S_z$-basis, with the *net* spin component $S_Z = 0$ (capital letters will denote the net value of an observable in question). When entangled, neither particle can be imaged as a single vector like in Fig.2 since neither of them has a definite state. We can only treat the system analytically.

A general expression for an entangled state with $S_Z = 0$ is

$$|\Psi\rangle_{AB} = p\,|\uparrow\rangle_A |\downarrow\rangle_B + \tilde{q}\,|\downarrow\rangle_A |\uparrow\rangle_B \quad \text{with } \tilde{q} \equiv q e^{i\alpha} \text{ and } p^2 + q^2 = 1 \quad (2.1)$$

Phase $\alpha$ here may include $\phi \equiv \varphi - \varphi'$. We call a case with opposite individual outcomes for measured observables $(-)$ correlations", and case with equal outcomes "$(+)$ correlations". In case (2.1), both outcomes $|\uparrow\rangle_A |\downarrow\rangle_B$ or $|\downarrow\rangle_A |\uparrow\rangle_B$ are $(-)$ correlations. Their relative weight and thereby entanglement strength is given by the ratio

$$\varepsilon \equiv p^2 / q^2 \quad (2.2)$$

The corresponding probabilities $\mathcal{P}_p$, $\mathcal{P}_q$ expressed in terms of $\varepsilon$ are

$$\mathcal{P}_p \equiv p^2 = \frac{\varepsilon}{1+\varepsilon}\,; \quad \mathcal{P}_q \equiv q^2 = \frac{1}{1+\varepsilon} \quad (2.3)$$

At weak entanglement ($\varepsilon \ll 1$ or $\varepsilon \gg 1$) one of the superposed outcomes in (2.1) becomes much more probable than the other, making the respective term overwhelmingly dominating and thus bringing *each* particle closer to the respective definite state. In the limit $\varepsilon \to 0$ or $\varepsilon \to \infty$, (2.1) reduces to only one term, the entanglement vanishes, each particle acquires its own state, but these individual states remain strictly $(-)$ correlated in the $S_z$-basis.

As with a single qubit, we can monitor the amplitudes of state (2.1) by changing the basis. Switch to basis $S_\mathbf{e}$ along some direction **e** as used in Sec.1. Writing (1.12) for A and then for B, using the rule $\theta' = \pi - \theta$ and putting into (2.1) gives after some algebra

$$|\Psi\rangle_{AB} = e^{-i\delta}\left[\tilde{f}\left(|\mathbf{e}\rangle_A |\mathbf{e}\rangle_B - |\bar{\mathbf{e}}\rangle_A |\bar{\mathbf{e}}\rangle_B\right) - \tilde{g}|\mathbf{e}\rangle_A |\bar{\mathbf{e}}\rangle_B - \tilde{h}|\bar{\mathbf{e}}\rangle_A |\mathbf{e}\rangle_B\right] = $$
$$= e^{-i\delta}\left[|\mathbf{e}\rangle_A \left(\tilde{f}|\mathbf{e}\rangle_B - \tilde{g}|\bar{\mathbf{e}}\rangle_B\right) - |\bar{\mathbf{e}}\rangle_A \left(\tilde{h}|\mathbf{e}\rangle_B + \tilde{f}|\bar{\mathbf{e}}\rangle_B\right)\right] \quad (2.4)$$

where

$$\tilde{f} \equiv (p+\tilde{q})mn, \quad \tilde{g} \equiv pm^2 - \tilde{q}n^2, \quad \tilde{h} \equiv pn^2 - \tilde{q}m^2 \tag{2.5}$$

The immaterial factor $e^{-i\delta}$ in (2.4) represents the symmetry of result (2.4) with respect to rotation of **e** around the Z-axis.

We have here the terms representing *both* types of correlations. If Alice finds A in the $|e\rangle_A$-state, it does not determine an outcome of her partner's Bob measurement on B. The outcome $|e\rangle_A$ collapses B to a *superposition* $\tilde{f}|e\rangle_B - \tilde{g}|\bar{e}\rangle_B$ rather than just to $|\bar{e}\rangle_B$, so there is a chance $|\tilde{f}|^2$ for Bob to get $(+)$ correlated outcome $|e\rangle_B$. And if Alice gets the result $|\bar{e}\rangle_A$, then B collapses to *another superposition* $\tilde{h}|e\rangle_B + \tilde{f}|\bar{e}\rangle_B$ instead of just $|e\rangle_B$, so there is the same chance $|\tilde{f}|^2$ for Bob to find B also in state $|\bar{e}\rangle_B$. In either case, only probabilistic prediction can be made for measurement on B.

The basis-dependence of the initially pure $(-)$ or $(+)$ correlations and their probabilistic nature had inspired the ground-breaking discussions of possibility of superluminal signaling between separated locations [6-9]. Some closely related topics can be found, e.g., in [10-15].

As pointed out before, *switching to an* **e** *-basis* is equivalent to inserting AP and BS in the bi-photon case, but the corresponding effect here is reached without any insertions.

Denote the respective probabilities of $(+)$ correlations in (2.4) as $\mathcal{P}^+(\mathbf{e}, \mathbf{e})$, $\mathcal{P}^+(\bar{\mathbf{e}}, \bar{\mathbf{e}})$, and their sum as $\mathcal{P}^+$. Similarly, $\mathcal{P}^-(\mathbf{e}, \bar{\mathbf{e}})$, $\mathcal{P}^-(\bar{\mathbf{e}}, \mathbf{e})$ will stand for $(-)$ correlations, and $\mathcal{P}^-$ for their sum. We can farther reduce the number of variables here by introducing, apart from $\varepsilon \equiv p^2/q^2$, also the ratio $\sigma \equiv m^2/n^2 = \tan^{-2}\frac{\chi}{2}$. Then

$$p^2 = \frac{\varepsilon}{1+\varepsilon}, \quad q^2 = \frac{1}{1+\varepsilon}; \quad m^2 = \frac{\sigma}{1+\sigma}, \quad n^2 = \frac{1}{1+\sigma} \tag{2.6}$$

All probabilities are then calculated from (2.4, 5) in terms of $\varepsilon, \sigma$:

$$\mathcal{P}^+(\mathbf{e}, \mathbf{e}) = \mathcal{P}^+(\bar{\mathbf{e}}, \bar{\mathbf{e}}) = |\tilde{f}|^2 = \sigma \frac{\varepsilon + 1 + 2\sqrt{\varepsilon}\cos\alpha}{(1+\varepsilon)(1+\sigma)^2}, \quad \mathcal{P}^+ = 2\sigma \frac{\varepsilon + 1 + 2\sqrt{\varepsilon}\cos\alpha}{(1+\varepsilon)(1+\sigma)^2} \tag{2.7}$$

$$\mathcal{P}^-(\mathbf{e}, \bar{\mathbf{e}}) = |\tilde{g}|^2 = \frac{\varepsilon\sigma^2 + 1 - 2\sqrt{\varepsilon}\sigma\cos\alpha}{(1+\varepsilon)(1+\sigma)^2}; \quad \mathcal{P}^-(\bar{\mathbf{e}}, \mathbf{e}) = |\tilde{h}|^2 = \frac{\varepsilon + \sigma^2 - 2\sqrt{\varepsilon}\sigma\cos\alpha}{(1+\varepsilon)(1+\sigma)^2} \tag{2.8}$$

$$\mathcal{P}^- = (1+\sigma)^{-2}\left(1 + \sigma^2 - 4\frac{\sqrt{\varepsilon}\sigma}{1+\varepsilon}\cos\alpha\right) \tag{2.9}$$

It is easy to see that $\mathcal{P}_{Net} = \mathcal{P}^+ + \mathcal{P}^- = 1$. Unlike the bi-photon case [3], here the two $(-)$ correlations have different probabilities, whereas chances for both $(+)$ correlations are the same. The reason is that path-entangled bi-photon in [3] is, in contrast to (2.1), a superposition of $(+)$ correlated states.

The visibilities for (2.7), (2.9) are

$$\mathcal{V}^+(\varepsilon) = 2\frac{\sqrt{\varepsilon}}{1+\varepsilon} \quad \text{for } \mathcal{P}^+ \qquad (2.10)$$

and

$$\mathcal{V}^-(\sigma, \varepsilon) = \frac{4\sqrt{\varepsilon}\sigma}{(1+\varepsilon)(1+\sigma^2)} \quad \text{for } \mathcal{P}^- \qquad (2.11)$$

They are identical to results in [3] except for swapping between $\mathcal{P}^+$ and $\mathcal{P}^-$ and between the respective visibilities due to mentioned turnover of correlations for considered systems.

Now, in the same way as we did in [3] for a bi-photon, we calculate the *local* probabilities, say, for qubit A to wind up in $|\mathbf{e}\rangle_A$ or in $|\bar{\mathbf{e}}\rangle_A$ from the initial entangled state:

$$\mathcal{P}(\mathbf{e}) \equiv \mathcal{P}^+(\mathbf{e}, \mathbf{e}) + \mathcal{P}^-(\mathbf{e}, \bar{\mathbf{e}}) = \frac{\varepsilon\sigma + 1}{(1+\varepsilon)(1+\sigma)} \qquad (2.12)$$

$$\mathcal{P}(\bar{\mathbf{e}}) \equiv \mathcal{P}^+(\bar{\mathbf{e}}, \bar{\mathbf{e}}) + \mathcal{P}^-(\bar{\mathbf{e}}, \mathbf{e}) = \frac{\varepsilon + \sigma}{(1+\varepsilon)(1+\sigma)} \qquad (2.13)$$

These two expressions are the basic result of the work. Just as in case of a bi-photon, local probabilities turn out to be phase-independent regardless of the ES, so there is no self-interference within all range of ES, $0 < \varepsilon < \infty$. The local coherence can exist only in a totally disentangled system.

### 3. "All roads lead to Rome"

The comparison of both cases shows unusual mix of differences and similarities eventually leading to the same result. Few things are more different from each other than a massless boson and massive fermion. It might seem quite natural that the ways of their monitoring must also be quite different. But a closer look shows that the difference is restricted to physical operations, while their mathematical descriptions turned out to be the same. No surprise that it gives *the same* new result for both cases. The result can be called "Total Mutual Intolerance" of local coherence and entanglement". The corresponding effect is utterly unusual and conflicts with a rule that two intimately linked characteristics of a simple system must change concurrently under changing general conditions. Local and global coherence in a bipartite are probably the most closely linked characteristics. It is natural to expect that both should *change*, even if differently, under changing ES. Most natural expectation is that the zero local coherence under maximal ES should start to gradually increase when entanglement weakens. Weakening entanglement can give more and more room for the local coherence. But the analysis shows that in contrast with the global coherence that changes with $\varepsilon$, the local coherence remains zero in all domain of $\varepsilon$ except for a totally disentangled system.

Actual situation is more subtle. Disentanglement can be obtained in two different ways: direct measurement which is a discontinuous process; or gradually decreasing ES according to (2.2). It is only the first way that resurrects local coherence; the second way is, as mentioned in Sec.2, only abstract mathematical limit at $\varepsilon \to 0$ or $\varepsilon \to \infty$. In other words, the second way does not lead to disentanglement on the local level. Changing conditions affect only global coherence while its local counterpart remains intact. This violates the above formulated Concurrency Rule.

The basic factor that makes such violation non-paradoxical is the innately probabilistic nature of QM. In Special Relativity, it is this factor that makes instant reconfiguration of a measured nonlocal quantum state consistent with relativity of simultaneity [16]. This is even more relevant in our case, when the studied characteristics are probabilities rather than observables.

Eventually, with all spectrum of possibilities, all the ways lead to the same result in both systems. This result opens an interesting question whether the revealed effect works only for bi-qubits, be they photonic or fermionic, or also for entangled systems with multiple eigenstates.

## Conclusions

The proposed thought experiments open a possibility to study coherence transfer between local and global scale for bi- photons and bi-fermions at varying ES. The analysis shows the same result for both systems despite their drastic difference. It can be formulated as violation of the concurrency rule and the accompanying "mutual intolerance" between the global and local coherence in an entangled bipartite. In contrast with incompatible observables like momentum and position, whose expectation values can still coexist under tradeoff between their indeterminacies, there is no coexistence for local and global coherence. If confirmed, the effect might open some new venues in Quantum Physics and Quantum Information Theory.

## Acknowledgements

I am grateful to Art Hobson for stimulating discussions that had inspired this work. I also want to thank Nick Herbert and Anwar Shiekh for their constructive comments.

## References

1. J. G Rarity, P. R. Tapster, Experimental violation of Bell's inequality based on phase and momentum, *Phys. Rev. Let*. **64**, 2496 (1990)
2. Z. Y. Ou, X. Y. Zou, L. J. Wang, and L. Mandel, Observation of nonlocal interference in separated photon channels, *Phys. Rev. Let*. **65**, 321 (1990)
3. M. Fayngold, Generalized rules of coherence transfer from local to global scale, arXiv:1908.06185 [quant-ph]
4. Moses Fayngold, Vadim Fayngold, *Quantum Mechanics and Quantum Information*, Wiley-VCH, Weinheim, 2013
5. M. Fayngold, Multi-Faced Entanglement, arXiv: 1901.00374 [quant-ph]
6. N. Herbert, FLASH: a superluminal communicator based upon a new kind of quantum measurement, *Found. Phys*., 12, 1171 (1982)
7. W. K. Wootters, W. H. Zurek, A single quantum cannot be cloned, *Nature*, **299**, 802 (1982)
8. D. Dieks, Communication by EPR devices, *Phys. Let*. *A*, **92** (6), 271 (1982)
9. S. J. van Enk, No cloning and superluminal signaling, arXiv: quant-ph/9803030 (1998)

10. V. Jacques, E. Wu, F. Grosshans, F. Treussart, , P. Grangier, A. Aspect, and J-F. Roch,
    Delayed-choice test of quantum complementarity with interfering single photons,
    *Phys. Rev. Lett*. **100**, 220402 (2008)
11. A. Peruzzo, P. Shadbolt, N. Brunner, S. Popescu, J. L. O'Brien,
    A quantum delayed-choice experiment, *Science* **338**, 634 (2012)
12. F. Kaiser, T. Goudreau, P. Milman, D. B. Ostrowsky, S. Tanzilli,
    Entanglement-enabled delayed-choice experiment, *Science* **338**, 637 (2012)
13. Manabendra Nath Bera, TabishQureshi, Mohd Asad Siddiqui, and Arun Kumar Pati,
    Duality of Quantum Coherence and Path Distingushability, arXiv:1503.02990v2 [quant-ph]
14. M. A. Nielsen, I. L. Chuang, *Quantum Computation and Quantum Information*,
    Cambridge Univ. Press*.,* 2004
15. S. Haroche and J.-M. Raimond, *Exploring the Quantum*, Oxford Univ. Press, 2006
16. M. Fayngold, How the instant collapse of a spatially-extended quantum state is
    consistent with relativity of simultaneity,
    arXiv:1605.05242 [physics.gen-ph]; *Eur. J. Phys*., **37** (6), 2016